\documentclass{elsarticle}

\usepackage{lineno,hyperref}
\modulolinenumbers[5]
\usepackage[utf8]{inputenc}
\usepackage{amsmath}
\usepackage{amsthm}
\usepackage{amsfonts}
\usepackage{amssymb}
\usepackage{amsthm}
\usepackage{tikz}
\usepackage{pstricks}
\usepackage{algorithm}
\usepackage{algorithmic}
\usepackage{lscape}
\usepackage{graphicx}
\usepackage{subcaption}
\usepackage[title]{appendix}

\journal{ ArXiv}

\usepackage[utf8]{inputenc}
\usepackage{amsmath}
\usepackage{amsfonts}
\usepackage{amssymb}
\def\bSig\mathbf{\Sigma}

\usepackage{amsmath}
\usepackage{amsfonts}
\usepackage{amssymb}
\usepackage{color}
\usepackage{tikz}
\usepackage{multirow}
\usepackage{lscape}
\usepackage[a4paper]{geometry}

\usepackage{pstricks}
\usepackage{algorithm}
\usepackage{algorithmic}

\newcommand\encircle[1]{%
  \tikz[baseline=(X.base)]
    \node (X) [draw, shape=circle, inner sep=0] {\strut #1};}

\begin{document}
\begin{frontmatter}

\title{A Multi-State Conditional Logistic Regression Model for the Analysis of Animal Movement}

\author[1]{ Aurélien Nicosia\fnref{mycorrespondingauthor}}
\ead{aurelien.nicosia.1@ulaval.ca}
\author[1]{ Thierry Duchesne}
\author[1]{ Louis-Paul Rivest}
\author[2]{ Daniel Fortin}
\address[1]{D\'{e}partement de math\'{e}matiques et de statistique, Universit\'{e} Laval, Qu\'{e}bec, Canada}
\address[2]{D\'{e}partement de Biologie, Universit\'{e} Laval, Qu\'{e}bec, Canada}



\cortext[mycorrespondingauthor]{Corresponding author}

\begin{abstract}
A multi-state version of an animal movement analysis method based on conditional logistic regression, called Step Selection Function (SSF), is proposed. In ecology SSF is developed from a comparison between the observed location of an animal and randomly sampled locations at each time step. Interpretation of the parameters in the multi-state model and the impact of different sampling schemes for the random locations are discussed. We prove the equivalence between the new model and a random walk model on the plane. This equivalence allows one to use both pure movement and local discrete choice behaviors in identifying the model's hidden states. The new method is used to model the movement behavior of GPS-collared bison in Prince Albert National Park, Canada. The multi-state SSF successfully teases apart areas used to forage and to travel. The analysis thus provides valuable insights into how bison adjust their movement to habitat features, thereby revealing spatial determinants of functional connectivity in heterogeneous landscapes.   

\end{abstract}

\end{frontmatter}
\linenumbers


\nolinenumbers

\section{Introduction}

In animal ecology, being able to understand and model the movement of animals is fundamental \citep{Nathan2008}.
For example, animal behaviourists want to see to what extent animals have preferred movement directions or are attracted
towards several environmental targets, such as food-rich patches and previously visited locations (spatial memory effect)
\citep{Latombe2014}. The development of Global Positioning System (GPS) technology permits the collection of a large
amount of data on animal movement. This can be combined to data available from geographic information systems (GIS)
to investigate how the environment influences animal displacement. To achieve this goal, robust statistical techniques and
flexible animal movement models are required.
\\
\\
Discrete time models for animal movement are actively being developed and investigated \citep{Holyoak2008}. Because displacement in discrete time
can be characterized by the distance and the direction between two consecutive localizations, circular-linear processes can be used to model movement in 2D.
A basic model is the biased correlated random walk \citep[BCRW,][]{Turchin}; it predicts the next motion angle as a comprise between the current one (often called directional persistence) and the direction towards a specific target (also called directional bias). This type of model handles environmental targets through their directions. Unfortunately, it cannot account for the impact of local pixel characteristics on the selection process. This can be done using \textit{Step Selection Function} (SFF), introduced by \cite{Fortin2005}. The SSF is a discrete choice model that compares the local characteristics of pixels selected by the animal at each time step (cases) to control pixels that could have been visited given the animal's previous position. 
\cite{Dancose2011}, \cite{Latombe2014} or \cite{Avgar2016} estimate the parameters of a BCRW and of the local selection probabilities using a single SSF while \cite{Duchesne2015} formally prove that the parameters of a BCRW
 can be estimated using a SSF.
\\
\\
Often, animal movement involves multiple states or behaviors \citep{Fryxell2008}. For instance, \cite{Langrock2012} identified two states, ‘‘exploratory’’ and ‘‘encamped’’, in their analysis of bison movement. The former state is characterized by long traveled distances and turning angles between two consecutive locations that tend to be concentrated around zero, while the latter  is characterized by short distances and nearly uniformly distributed turning angles. Multiple movement behaviors can be accounted for through hidden states. \cite{baum66} give a general presentation of these models and \cite{Morales2004}, \cite{Jonsen2005}, \cite{Holzmann2006}, \cite{Langrock2012} and more recently \cite{Nicosia16}  use hidden state models to analyze angular-distance data on animal movement
\\
\\
 The main contribution of this paper is, in Section 2, to propose a multi-state SSF model handling both animal movement and local habitat selection. Section 2 also discusses parameter interpretation in a multi-state context, and the sampling of control locations. In Section 3, we prove that the proposed multi-state SSF model can be used to fit the  multi-state random walk model of \cite{Nicosia16}; this theoretical result is validated using a simulation study and the analysis of a real data set. Section 4 identifies two states in the analysis of the movement trajectory of bison in Prince Albert National Park, Canada.  Both movement and local habitat selection parameters can vary between states.


\section{Multi-state Step Selection Function}
\label{s:MultiSSF}

\subsection{Single-state Step Selection Function}

Let us suppose that we follow an animal equipped with a GPS collar which provides the animal's location at regular time intervals $t=1,\ldots,T$, for example every 1 hour. The  data are combined with information on the animal's habitat in a geographic information system (GIS).  Step Selection Functions \citep[SSF,][]{Fortin2005}, specify how the animal uses its habitat by modeling the discrete choices that it makes at each time step. At a given time step, an SSF compares the characteristics of the location (and of the trajectory leading to this location) visited by the animal  with $J$ control steps to other locations that the animal could have visited at that time given the previous step. This comparison uses GIS data, an $r \times 1$ vector $\mathbf{x}_{0t}$, for the observed location and the corresponding vectors $\mathbf{x}_{jt}, j=1,\ldots,J$ for the $J$ control locations.

The step characteristics $x$ that comprise $\mathbf{x}_{jt}$  are of several types.  
An SSF compares the characteristics of observed steps and control steps. Steps can be characterized by the features that can be encountered along the step (e.g., road, proportion of forest cover), at the end of the step (e.g., particular land cover type, elevation), bearing direction with respect to habitat features at relatively far distances (e.g., road, canopy gap) and speed (\cite{Fortin2005}, \cite{Dancose2011}, \cite{Vanak2013}, \cite{Latombe2014},    \cite{Basille2015}).
The wish of an animal to go to a specific location, e. g. a target meadow, can be entered in the model as an explanatory variable equal to the cosine of the difference between the direction to the next location and the direction to the target \citep{Dancose2011}. A directional persistence, that is the wish of an animal to move forward, can be included in the analysis through the cosine of the difference between the motion angles of current and previous steps. \cite{Latombe2014} show, using data on caribou, how we can include different types of characteristics in SSF models.

The data for an SSF analysis is $\{[\mathbf{x}_{0t},\mathbf{x}_{1t},\ldots,\mathbf{x}_{Jt}]:  t=1,\ldots T\}$. It is analyzed using a conditional logistic regression model for a  matched case-control design (\cite{Hosmer}, chapter 7). It is also equivalent to the multinomial logit discrete choice model \citep[][chapter 3]{train2003discrete}. Thus, at time step $t$, the probability that the animal chooses the location with step characteristics $\mathbf{x}_{0t}$, rather than one of the $J$ control locations with respective step characteristics $\mathbf{x}_{jt}, j=1,\ldots,J$ is
\begin{equation}\label{probaChoix}
p_{t}=\frac{\exp \left(\mathbf{x}_{0t}^\top \beta\right)}{\sum_{j = 0}^J \exp \left(\mathbf{x}_{jt}^\top \beta\right)},
\end{equation}
where $\beta$ is a $ r \times 1$ vector of unknown selection parameters.
Following \cite{Hosmer}, $\beta$ is easily estimated by maximizing the conditional logistic regression likelihood given by
\begin{equation} \label{LikeK1}
L(\beta)=\prod_{t=1}^T \frac{\exp \left(\mathbf{x}_{0t}^\top \beta\right)}{\sum_{j = 0}^J \exp \left(\mathbf{x}_{jt}^\top \beta\right)}.
\end{equation}

To discuss the interpretation of $\beta$ we first consider  a simple model with a single dichotomous explanatory variable $x$ identifying a particular type of habitat representing $100\times H$\% of the study area. In a null model, with $\beta=0$, the probability of selecting the habitat at a time step is $H$.  When $\beta \ne 0$ this probability becomes $e^{\beta} H/(1+e^{\beta} H) $  which is larger than $H$ if $\beta >0$, see Appendix of \cite{Duchesne2015} for more details. With a continuous explanatory variable, the same interpretation holds. Suppose that a variable $x$, available at each location of the map, is distributed as a stationary random field with marginal density $f(x)$. If $\beta = 0$, then the density of $x$ for the selected locations is $f(x)$.  When $ \beta \neq 0 $, this density is proportional to $e^{\beta x} f(x)$; this gives a weighted  distribution (see, \cite{Patil2005}). If $\beta>0$ the animal tends to select locations with higher values of $x$ more often than would be expected with a purely random selection. As a matter of fact, the value of $\beta$ is actually the log of the odds that the animal will choose a location with a value of the explanatory variable equal to $x+1$ divided by the odds of choosing a location with a value of the explanatory variable equal to $x$.

When the $x$ variable is the cosine of the difference between the angles of the direction of a potential target and the current motion angle, then a positive value of $\beta$ means that
the target is attractive (steps in its direction are selected more often). More details on this latter interpretation may be found in \cite{Duchesne2015} who actually show that a SSF with such
 cosine explanatory variables and uniform sampling of the control locations is equivalent to a  Biased Correlated Random Walk model (BCRW, \cite{Turchin}).

\subsection{Extension to multi-state SSF}
Often, the animals exhibit more than a single step selection behavior \citep{Fryxell2008}. Such a change in behavior can be explained by a hidden-state model \citep{Sylvia13} with a different SSF in each state. Let $\beta^{(k)},k=1,\ldots,K$ denote the selection coefficients of the SSF when the animal is in state $k$. To model the animal's unobserved behavioral state over time, we consider a hidden process $\{{S}_t,t=1,\dots,T\}$ where the value of $S_t$ represents the state (behavior) in which the animal is at time step $t$. Following the reasoning of \cite{Nicosia16}, the likelihood function of this multi-state SSF model is

\begin{equation} \label{Like}
L(\mathbf{\beta})=\prod_{t=1}^T \sum_{k=1}^K \left( p_t^{(k)}   \cdot \mathbb{P}(S_t=k|\mathcal{F}_{t-1}^c) \right),
\end{equation}
where
\begin{equation} \label{ptk}
p_t^{(k)} = \frac{\exp \left(\mathbf{x}_{0t}^\top \beta^{(k)}\right)}{\sum_{j = 0}^J \exp \left(\mathbf{x}_{jt}^\top \beta^{(k)}\right)},
\end{equation}
 and $\mathcal{F}_t^c$ denotes the complete data history up to time $t$, which consists of the observed data and of the unobserved state; thus $\mathcal{F}_t^c$ contains ${S}_\ell$ and $[\mathbf{x}_{0\ell},\mathbf{x}_{1\ell},\ldots,\mathbf{x}_{J\ell}]$, for  $\ell=1,\ldots, t$.  Figure~\ref{f:dependence} presents the dependence structure of the proposed model.
\begin{figure}[H]
\centering
$
\begin{array}{cccccc}
\textbf{Hidden state:}   & \longrightarrow & \encircle{$S_{t-1}$} & \longrightarrow & \encircle{$S_{t}$} & \longrightarrow   \\
  & & \downarrow & & {\downarrow}&    \\
\textbf{Observed choice}   &  &   \boxed{p_{t-1}^{(k)}} &  & \boxed{p_t^{(k)}} &
  \\
 &  & \uparrow & \searrow& {\uparrow}&     \\
\textbf{Information:}   &  &   {\mathcal{F}_{t-2}^o} &  & {\mathcal{F}_{t-1}^o}   &
\end{array}
$
\caption{Dependence structure of the proposed model}\label{f:dependence}

\end{figure}

 The probability $\mathbb{P}(S_t=k|\mathcal{F}_{t-1}^c)$ in (\ref{Like}) is called the ``predictive'' probability. It can be efficiently computed using a filtering-smoothing algorithm (see \ref{a:D} for more details) when $\{S_t\}$ is modeled as a Markov chain.

Inference about the parameters $\beta^{(k)},k=1,\ldots,K$ is based on the maximized log-likelihood $\ell(\mathbf{\beta})=\ln L(\mathbf{\beta})$. When $\{S_t\}$ is modeled as a Markov chain, we can use the EM algorithm and the filtering-smoothing algorithm to implement the inference and standard errors are estimated by computing the hessian of the observed log-likelihood function. Details of the procedure are given in \ref{a:C}. To reach the global maximum of the observed likelihood function we use the short-run long-run EM algorithm strategy, see Appendix C of \cite{Nicosia16} for more details.
\subsubsection{Interpretation of the parameters}
The interpretation of a multi-state SSF parameter $\beta^{(k)}$ is almost the same as that of $\beta$ in a single-state SSF, except that it is conditional on the state in which the animal
is.  A non null $\beta$ for a variable $x$ means that the distribution of $x$ constructed with the locations chosen by the animal differs from the stationary distribution of $x$, $f(x)$ over the study area. For instance if we have two states ($k=1,2$) and we have a coefficient $\beta^{(1)}>0$ for $x$ in state 1 and a coefficient $\beta^{(2)}<0$ for $x$ in state 2, then this means that when the animal is in state 1, it tends to select locations, with distribution proportional to $e^{\beta^{(1)}x} f(x) $, with high values of $x$ more often while in state 2 it tends to favor locations, with distribution proportional to $e^{\beta^{(2)}x} f(x) $, with small values of $x$. The value of $\beta$ has the same
log odds ratio interpretation within each state as in the single-state SSF.

\subsubsection{Sampling the control locations}
\label{s:SamplingProcedure}

This section discusses the sampling of the control locations. At time $t$ the animal is at a location  $P_t\in \mathbb{R}^2$ and at time $t+1$ he will be at location $P_{t+1}$.  The control locations for $P_{t+1}$ are defined as a direction and a distance from $P_t$. Following \cite{Duchesne2015} we select the control directions uniformly on $[0,2\pi[$. \cite{Forester2009} argue that
the method used to select the control distances influences the parameter estimates in a standard, single state, SSF. They emphasized that the range of the control distances needs to cover all the distances that the animal may possibly travel.

 Let $\{(\phi_{jt},h_{jt}):\ j=1,\ldots,J\}$, where $\phi_{jt}\in [0,2 \pi[$ is an angle and  $h_{jt}>0$ is a distance, be the polar coordinates (with $P_t$ as origin) of the $J$ control locations matched with $P_{t+1}$, and $(\phi_{0t},h_{0t})$ be the polar coordinates of $P_{t+1}$. As discussed above, $\phi_{jt}$ is sampled uniformly over $[0,2 \pi[$. The distances can be sampled uniformly over $[0,M]$, where $M$ is large enough for $[0,M]$ to cover all possible observed distances. Let $D_{k},k=1,\ldots,K$, denote the support of the traveled distances in state $k$. \cite{Forester2009} have shown that, in a single state model, if the support of the control distances (i.e., $[0,M]$) does not include the support $D_1$ of the traveled distances, then we induce a bias in the estimation of the parameters $\beta^{(1)}$. This statement also applies to a multi-state model and therefore $[0,M]$ needs to cover $\cup_k D_k$.

 Another way to sample the control distances is through a parametric distribution, such as a gamma distribution. This latter sampling procedure is discussed by \cite{Forester2009} for a single state SSF. We implement it in a multi-state setting in the next section.

\section{Multi-State SSF Model with Distances and Angles}
\label{s:global}
\subsection{Equivalence with a Random Walk Model} \label{s:equivalence}
In this section we investigate whether the multi-state BCRW (Biased Correlated Random Walk) model introduced by \cite{Nicosia16} can be fitted using a multi-state SSF. Our goal is to generalize the findings of \cite{Duchesne2015} to a complex multi-state SSF involving state dependent distributions for distances. This highlights that the parameters of a multi-state SSF can be interpreted as those of a multi-state BCRW.

\begin{figure}[H] 
\centering
\begin{tikzpicture}[scale=0.8]

\draw  (-1,-1) node[below]{$P_{t-1}$} node{$\bullet$}; 
\draw (1,1) node[below right]{$P_{t}$} node{$\bullet$}; 
\draw [->] (-1,-1)--(0.95,0.95);
\draw [dashed] (-1,-1) -| (1,-1);
\draw (0,-1) arc (0:45:1) ;
\draw (0.5,-0.5) node {$\phi_{0,t-1}$};
\draw (-0.4,0.3) node {$h_{0,t-1}$};

\draw (9.8,-0.5) node[below right]{} node{Target 1};
 
\draw[] [thick,dotted,->] (1,1) -- (8.9,-0.45) ;
\draw (4,1) arc (0:-33:1) ;
\draw[] (4,0.6) node[right] {$\theta_{1,t}$};

\draw (2,1) arc (0:100:1) ;
\draw[] (1.4,2) node[right] {$\theta_{2,t}$};

\draw (0,6.2) node[right]{} node{Target 2}; 
\draw[] [thick,dotted,->] (1,1) -- (0,5.9);

\draw (9.7,-0.4) node[below right]{} ;
\draw (0.7,6) node[right]{};

\draw[] (8,4) node[below right]{$P_{t+1}$} node{$\bullet$}; 
\draw (2.3,3) node[below right]{} ;
\draw[] [->] (1,1)--(7.9,3.95);

\draw (3,1) arc (0:42:1) ;
\draw[] (3.1,1.5) node[right] {$\phi_{0,t}$};
\draw [dashed] (1,1) -| (5,1);
\draw[] (4,2.7) node {$h_{0,t}$};

\end{tikzpicture}
\caption{Notation of BCRW model with $p=2$ targets.}\label{graphNotation}
\end{figure}
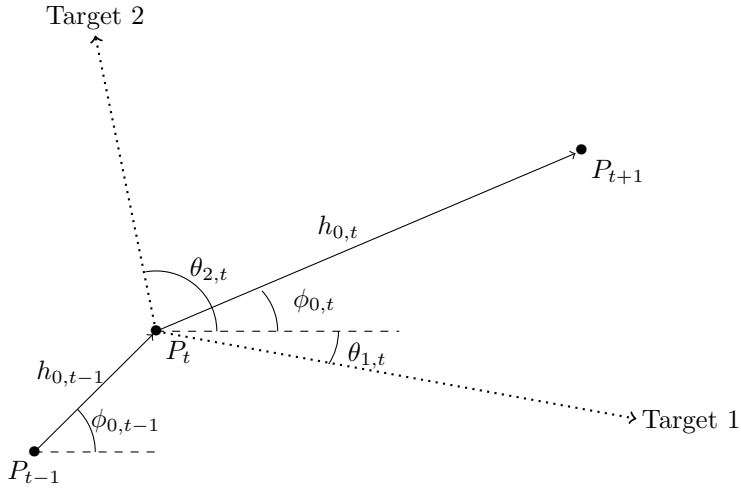

To see this we let both models depend on the directions from $P_{t-1}$ to $P_{t}$, $\phi_{0,t-1}$, and from $P_{t}$ to $p$ potential targets in the landscape (e.g. the closest meadow, a canopy gap or the closest forest), denoted by the angles $\theta_{it}, i =1,\ldots,p$. Figure~\ref{graphNotation} exposes the notations with $p=2$ targets. 

Knowing that the animal is in state $k$,  The distribution of the direction $\phi_{0,t}$ at time $t$, observed when traveling  from $P_t$ to $P_{t+1}$, depends on the vector
\begin{equation}\label{V}
\mathbf{V}_t^{(k)}=\kappa_0^{(k)} \left(\begin{array}{c}
\cos(\phi_{0,t-1})\\
\sin(\phi_{0,t-1})
\end{array}  \right)+\sum_{i=1}^p \kappa_i^{(k)} \left(\begin{array}{c}
\cos(\theta_{it})\\
\sin(\theta_{it})
\end{array} \ \right) , t=1,\ldots,T,
\end{equation}
where $(\kappa_0^{(k)},\dots,\kappa_p^{(k)})$, $k=1,\ldots,K$ are unknown parameters depending on the state $k$. The direction $\phi_{0,t}$ is assumed to have a von Mises distribution \citep[see,][]{Mardia00} that depends on state $k$.
The mean direction is the direction of $\mathbf{V}_t^{(k)}$ and the concentration parameter is the length of $\mathbf{V}_t^{(k)}$, which correspond to the consensus model proposed by \cite{Rivest2015}.

The traveled distances for the BCRW in state $k$, are assumed to follow a distribution with density from the following exponential family (see \cite{lehmann2003theory}, section 1.5):
\begin{equation} \label{distexpo}
g_k(d; \eta^{(k)})= b(d) \exp \lbrace \eta^{(k)\top} T(d)-A(\eta^{(k)}) \rbrace, d>0, k=1,\ldots,K .
\end{equation}
In (\ref{distexpo}) $\eta^{(k)} \in \mathbb{R}^m$  is the vector of natural parameters, $T$ is a $\mathbb{R}^m$ valued vector of sufficient statistics, $b$ is a positive function, and $A$ is a $\mathbb{R}$ valued function called the log-partition function.

A SSF that is equivalent to the BCRW specified by (\ref{V}) and (\ref{distexpo}) has covariates that depend on the sufficient statistic $\{T(h_{jt})\}$  in (\ref{distexpo}) and the cosines of the differences between $\phi_{jt}$ and the directions to  potential targets, $(\cos(\phi_{jt}-\phi_{0,t-1}),\cos(\phi_{jt}-\theta_{1t}),\dots,\cos(\phi_{jt}-\theta_{pt}))^\top$.
Thus the vector of explanatory variables for the SSF is
\begin{equation} \label{explanatorySSF}
\mathbf{x}_{it}=\left(T(h_{it}),\cos(\phi_{it}-\phi_{0,t-1}),\cos(\phi_{it}-\theta_{1t}),\dots,\cos(\phi_{it}-\theta_{pt})  \right)^\top.
\end{equation}

With this specification of $\mathbf{x}_{it}$, one has $\beta^{(k)}=(\eta^{(k)}, \kappa_0^{(k)}, \kappa_1^{(k)}, \ldots, \kappa_p^{(k)})$
and the function $b$ in (\ref{distexpo}) appears in the SSF model as the offset $\log(b)$.
The numerator of $p_{t}^{(k)}$ in (\ref{ptk}) becomes
\begin{equation}\label{numSSFBCRW}
e^{\mathbf{x}_{0t}^\top {\beta}^{(k)}+\ln b(h_{0t})}=  b(h_{0t})e^{\eta^{(k)} T(h_{0t})} \times e^{ \kappa_0^{(k)} \cos(\phi_{0t}-\phi_{0,t-1}) +\sum_{i=1}^p \kappa_i^{(k)} \cos(\phi_{0t}-\theta_{it})},
\end{equation}
which is the product of two terms, one for the distances and one for the directions. Note that (\ref{numSSFBCRW}) is the numerator of the time $t$ contribution to the conditional likelihood function, for a given state, of the multi-state BCRW of \cite{Nicosia16}.

The denominator of $p^{(k)}_t$ is proportional to $J^{-1}\sum_{j=0}^J\exp(\mathbf{x}_{jt}^\top {\beta}^{(k)}+\log b(h_{jt}))$. The limit of this denominator as $J$ goes to infinity depends on the way in which the controls have been selected. As mentioned in Section \ref{s:SamplingProcedure}, the controls angles $\{\phi_{jt}\}$ are drawn using a uniform distribution over $[0, 2 \pi[$. One can sample the control distances uniformly in $[0,M]$, with a large $M$ value as recommended in Section \ref{s:SamplingProcedure}. With these methods for selecting the controls, the denominator is approximatively equal to the denominator of the time $t$ contribution to conditional likelihood, for a given state, of the multi-state BCRW of \cite{Nicosia16}. Details are given in \ref{a:B}

Let us now suppose that the control distances are sampled from (\ref{distexpo}), with a vector of parameters $\tilde{\eta}$.
Note that because the states are unobserved, the distribution from which the control locations are sampled cannot depend on the state, and hence $\tilde{\eta}$ is constant over $k$.
In this case the offset does not appear anymore in $p^{(k)}_t$ as it is included in the density of the control distances that are sampled according to (\ref{distexpo}). The offset is needed when the distances are sampled uniformly since $b(d)$ does not appear in the density of the control locations for this sampling scheme. Using the weak law of large numbers, the denominator of $p^{(k)}_t$ is then approximatively equal to a tilted version of the denominator of the time $t$ contribution to
conditional likelihood, for a given state, of the multi-state BCRW  of \cite{Nicosia16}. Thus in (\ref{distexpo}), the parameter $\eta^{(k)}$ is replaced by $\eta^{(k)}+\tilde{\eta}$. Thus if $\hat{\eta}_{\text{SSF}}^{(k)}$ is the SSF estimator for $\eta^{(k)}$, then the corresponding BCRW estimator is
\begin{equation} \label{biascorrection}
\hat{\eta}^{(k)} = \hat{\eta}_{\text{SSF}}^{(k)}+\tilde{\eta}.
\end{equation}
A detailed proof is provided in \ref{a:B}.

%

\subsection{Simulation Studies}
\label{s:simul}
In Section \ref{s:equivalence} we established the equivalence between the multi-state SSF and the multi-state BCRW models when the number of control locations sampled $J$ was large
using the law of large numbers. We now investigate whether this equivalence holds for a finite value of $J$ with both the uniform and parametric sampling schemes for the control distances.
We also assess the adequacy of the bias correction (\ref{biascorrection}) for the estimators of the distance coefficients proposed in the case of parametric sampling.
We follow the simulation studies of \cite{Nicosia16} which investigate the statistical properties of a general BCRW model.
In this simulation study we intend to demonstrate that if we simulate a trajectory from the general multi-state BCRW model then we can estimate its parameters using the proposed multi-state SSF model of Section~\ref{s:MultiSSF}.

We simulated the movement of one animal in the plane.
The simulation procedure includes one target and it was placed at the center of the map and the covariate $\theta_t$ represents the direction from the animal at position $P_t$ to this target at time step $t$. The simulation scenario consisted in repeating the following steps  $N=500$ times: (i) a time-homogeneous two-state Markov chain $S_{0:T}$ with transition matrix ${\cal{P}}$ is generated; (ii) at time 0, the animal is placed at a random position close to the south west corner of the map; (iii)
at each time step $t$, $t=1,2,\ldots$, the location of the animal is obtained by simulating a direction $\phi_{t0}$ and a distance $h_{t0}$ from the proposed general random walk model of Section \ref{s:equivalence} with $\phi_{0t}$ generated according to a consensus von Mises model with parameters $\kappa_0^{(k)},\kappa_1^{(k)}$, $k=1,2$ and explanatory angles $\phi_{0,t-1}$ and $\theta_t$ and $h_{0t}$ is simulated from a gamma distribution with shape parameter $\lambda_1^{(k)}$ and rate parameter  $1/\lambda_2^{(k)}$; (iv) the simulation stops when the animal is within 30 distance units from the target.
The gamma distribution belongs to the exponential family (\ref{distexpo}) with sufficient statistics $T(d)= (\log(d), -d)$ and vector of natural parameters $\eta^{(k)}= \left( \lambda_1^{(k)}-1 , 1/\lambda_2^{(k)}\right)^T$
 since its density can be written as
$$
 f(d;\lambda_1^{(k)},\lambda_2^{(k)})=\exp{ \left\lbrace  \left( \begin{array}{cc}
 \log(d) & -d
 \end{array}  \right) \left(\begin{array}{c}
 \lambda_1^{(k)}-1 \\
 1/\lambda_2^{(k)}
 \end{array}  \right)   -A( \lambda_1^{(k)}, \lambda_2^{(k)}) \right\rbrace},
$$
where $A( \lambda_1^{(k)}, \lambda_2^{(k)}) = \log \left(\Gamma(\lambda_1^{(k)})  \right) +\lambda_1^{(k)} \log(\lambda_2^{(k)}) $.

The values of the parameters used in the simulations are given in Table~\ref{t:scenario}.
The scenario is one where the animal shows high directional persistence and high attraction to the target when in state 1, and high directional persistence and a moderate repulsion from the target in state 2.
\begin{table}[H]
\caption{Parameters for the simulation scenario}
\centering
\begin{tabular}{c c c  }
\hline

 &\\
 &${\cal{P}}= \left( \begin{array}{cc}
 1-q_{1} & q_{1} \\
 q_{2} & 1-q_{2}
 \end{array} \right)$  &  $\left( \begin{array}{cc}
 0.9 & 0.1 \\
 0.2 & 0.8
 \end{array} \right)$  \\
 &     &   \\
 & $\kappa_0^{(1)}=\beta_3^{(1)}$  & 20   \\
 & $\kappa_1^{(1)}=\beta_4^{(1)}$  & 15   \\
&$\lambda_1^{(1)} - 1=\beta_1^{(1)}$  &4  \\
&$1/\lambda_2^{(1)}=\beta_2^{(1)}$  &10/7   \\

 &     & \\
&$\kappa_0^{(2)}=\beta_3^{(2)}$ & 10  \\
  &$\kappa_1^{(2)}=\beta_4^{(2)}$    & -2  \\
&$\lambda_1^{(2)}-1=\beta_1^{(2)}$ & 0 \\
&$1/\lambda_2^{(2)}=\beta_2^{(2)}$ & 2 \\
\hline
\hline
\end{tabular}
\label{t:scenario}
\end{table}
\noindent
with $q_1$ the probability $\mathbb{P}(S_t=2 | S_{t-1}=1)$ that the animal switches from state 1 at time $t-1$ to state 2 at time $t$. Similarly, $q_2$ denotes $\mathbb{P}(S_t=1 | S_{t-1}=2)$.

Once an animal's trajectory has been simulated, two sets of $J=500$ control locations, for each visited location, are sampled. In the first one, the control distances are sampled uniformly over $[0,15]$, where $M=15$ is large enough to cover the supports of the gamma distributions in the two states up to their 99.9th percentiles.  In the second set, the control distances are sampled from (\ref{distexpo}) with parameter $\tilde{\eta} = (0, 1)$, which actually corresponds to the exponential distribution with rate 1. The correspondence equation (\ref{biascorrection}) for the SSF estimators $(\widehat{\eta_{1,SSF}^{(k)}}, \widehat{\eta_{2,SSF}^{(k)}})= \left(\widehat{\left(\lambda_{1,SSF}^{(k)}-1\right)}, \widehat{1/\lambda_{2,SSF}^{(k)}}\right)$  and the parameters of model (\ref{distexpo}) are:
\begin{eqnarray*}
 \widehat{\eta_{1}^{(k)}} &= & \widehat{\eta_{1,SSF}^{(k)}}, k=1, 2. \\
 \widehat{\eta_{2}^{(k)}} &= & \widehat{\eta_{2,SSF}^{(k)}} +1 , k=1, 2.
\end{eqnarray*}

For this model the covariates (\ref{explanatorySSF}) are
$$
\mathbf{x}_{it}=\left(\log(h_{it}),-h_{it},\cos(\phi_{it}-\phi_{0,t-1}),\cos(\phi_{it}-\theta_{t})\right)^\top.
$$
 Note that here the offset is $\ln b(h_{0t})=\ln 1=0$.
With this definition of $\mathbf{x}_{it}$, the parameters of the SSF are $(\beta_1^{(k)}, \beta_2^{(k)}, \beta_3^{(k)}, \beta_4^{(k)})=(\lambda_1^{(k)}-1, 1/\lambda_2^{(k)}, \kappa_0^{(k)}, \kappa_1^{(k)})$ for $k=1,2$.

To evaluate the sampling properties of the SSF estimators, the following statistical indicators were calculated:
\begin{eqnarray}
b(\hat{\beta})&=& \frac{1}{500} \sum_{i=1}^{500}(\hat{\beta}^{(i)}-\beta) \label{biais}\\
\text{Sd}(\hat{\beta})&=& \sqrt{\frac{1}{499} \sum_{i=1}^{500}(\hat{\beta}^{(i)}-\bar{\beta})^2 } \label{RMSE},
\end{eqnarray}
where $\beta$ is a true parameter presented in Table~\ref{t:scenario}, $\hat{\beta}^{(i)}$ the parameter estimate in the $i$-th simulation and $\bar{\beta}$ the mean of the estimates over the 500 simulations. Equation (\ref{biais}) gives the bias of the estimator, (\ref{RMSE}) its standard deviation. When the control distances were generated from the unit exponential, the bias correction proposed in Section \ref{s:equivalence} was applied.

 \begin{table}[h!]
 \caption{Result of the $N=500$ simulations with $J=20,500$. The bias $b(\hat{\beta})$ at precision $10^{-3}$ is presented and its standard deviation between parenthesis (Sd$(\hat{\beta})$). }
 \centering
 \label{tab:resultSimul}
\begin{tabular}{|c c c c c |}
\hline
Multi-state BCRW &   \multicolumn{4}{c|}{multi-state SSF estimation}  \\
&   \multicolumn{2}{c}{Uniform sampling} &   \multicolumn{2}{c|}{Parametric (\ref{distexpo}) sampling} \\

True parameter & $J=20$ & $J=500$  & $J=20$ & $J=500$\\
\hline
\hline
$q_1=0.1$ &  -0.015 (0.03)& -0.000 (0.02)  & 0.013 (0.02) &0.002 (0.02)  \\
$q_2=0.2$ &-0.028 (0.03) & 0.002 (0.03) &-0.010 (0.04) &0.007 (0.03) \\
 & & &   &  \\
$\kappa_0^{(1)}=\beta_3^{(1)}=20$ &2.160 (3.19) & 0.177 (1.70) &  1.682 (5.47)  &0.305 (1.73)  \\

$\kappa_1^{(1)}=\beta_4^{(1)}=15$  & 2.291(2.67)& 0.212 (1.33)  & 0.644 (4.83)& 0.212 (1.29) \\

$\lambda_1^{(1)}-1=\beta_1^{(1)}=4$  & 0.031 (0.98)&    0.055 (0.43) & -1.691 (2.13) & 0.045 (0.47)\\

$1/\lambda_2^{(1)}=\beta_2^{(1)}=10/7$ & -0.008 (0.27)& 0.014 (0.13) & -2.701 (0.77) &0 .007 (0.15)\\
  & & &   &  \\

$\kappa_0^{(2)}=\beta_3^{(2)}=10$  & 1.575 (2.53)& 0.341 (1.27)  & 0.385 (1.66)&0.258 (1.11)\\

$\kappa_1^{(2)}=\beta_4^{(2)}=-2$  & 0.688 (1.16) &-0.050 (0.58)& -0.244 (0.87) & -0.072 (0.49) \\

$\lambda_1^{(2)}-1=\beta_1^{(2)}=0$  &-0.485 (0.35) &-0.006 (0.12)& -0.148 (0.23)&0.010 (0.10) \\

$1/\lambda_2^{(1)}=\beta_2^{(2)}=2$  &-1.046 (0.53)&  -0.040 (0.32) & -2.346 (0.80)&  0.062 (0.29)\\
\hline
\end{tabular}
 \end{table}

Table~\ref{tab:resultSimul} shows that the SSF recovers well the corresponding parameters of the BCRW model.
Under uniform sampling of the control distances, the bias in the estimators is negligible. When the control distances are sampled
from the unit exponential distribution,  (\ref{biascorrection}) gives consistent estimator of the distance parameters. 
Table~\ref{tab:resultSimul} exposes that the SSF does not recover well the corresponding parameters of the BCRW model when the number of control locations $J$ is small ($J=20$). The reason is that the equivalence of both models is established under the law of large numbers which assumes that the $J$ control locations have to be large enough. The simulation study highlights that SSF estimates are consistent for the parameters of the underlying BCRW model specified by (\ref{V}) and (\ref{distexpo}).


An additional illustration of the equivalence of the two methods, either SSF or BCRW, to fit a multi-state BCRW is presented in \ref{a:A}. It fits a BCRW to bison movement data and obtains nearly identical estimates with the two estimation methods.

\section{Multi-state SSF model with animal movement and resource selection}
\label{s:analysis}

\begin{figure}[h!]

\centering
\includegraphics[scale=0.4]{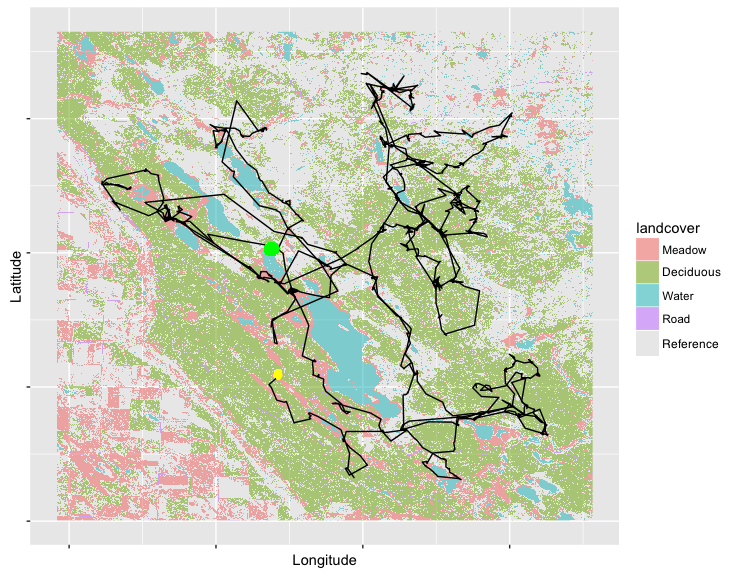}
\caption{Trajectory of a radio-collared bison from November 2013 to April 2014
in Prince Albert National Park. The green point is the departure and the yellow one the end. The landcover type `Reference' denotes the reference landscape type which mainly consists of a mixture of deciduous and conifer forests.}
\label{f:TrajecObsBison}
\end{figure}

\cite{Dancose2011} and \cite{Latombe2014} showed how a single state SSF can integrate both movement (angles and distances) and resource selection.  \cite{Avgar2016} further studied the properties of the approach.
When a large number of control locations are sampled, we have shown that the multi-state random walk model is equivalent to the multi-state SSF model and thus our proposed model can be viewed as a multi-state version of these SSF models.
In this section we analyse the trajectory of an individual bison from November 2013 to April 2014 ($T=3073$ hourly steps) in Prince Albert National Park, Saskatchewan, Canada. We show that two states can be distinguished and that the movement and selection parameters can vary betweeen states.  Figure~\ref{f:TrajecObsBison} depicts this trajectory.

During the winter season the bison tends to select more locations among meadows, water, roads or deciduous stands \citep{Dancose2011}. We can therefore treat all other types of landscape as
the baseline landscape level and fit the model with the following linear predictor:
\begin{eqnarray}
\mathbf{x}_{jt}^T \beta^{(k)} & = & \beta_{\text{cos.persis}}^{(k)} \cos(\phi_{jt}-\phi_{0,t-1})  \\
&+& \beta_{\text{dist.neg}}^{(k)} (-h_{jt})+\beta_{\text{dist.log }}^{(k)} \log(h_{jt}) \nonumber \\
& +& \beta_{\text{water}}^{(k)} z_{jt,\text{water}} +\beta_{\text{dec}}^{(k)} z_{jt,\text{dec}} \nonumber \\
&+& \beta_{\text{meadow}}^{(k)} z_{jt,\text{meadow}}+\beta_{\text{road}}^{(k)} z_{jt,\text{road}},
\ \ k=1,2  \nonumber
\label{eq:modelsect4}
\end{eqnarray}
$j=0,\ldots,J$, where the explanatory variable $z_{jt,*}$ is the indicator that the location $j$ at time step $t$ is of type *, with * denoting one of the four types of landscape; for example $z_{jt,\text{water}}=1$ if the $j$-th sampled location at time step $t$ is in the water and 0 otherwise. The $K=2$ states was validated by exploratory analyses similar to those presented by \cite{Nicosia16}.
Table~\ref{t:resulSSF} presents the estimates of the parameters by (\ref{eq:modelsect4}) along with their standard errors obtained when fitting the proposed SSF model with $J=500$ uniformly sampled control locations.


\begin{table}[h!]
\caption{Estimated parameters of the multi-state SSF model with $J=500$ uniformly control locations. The estimated parameters are presented and the standard errors are given between parentheses.}

\centering
\begin{tabular}{ccc}
Parameter &State 1: encamped & State 2: exploratory \\
  \hline
  \hline
  $q$ & 0.247 (0.02)& 0.161 (0.01)\\
$\beta_{\text{cos.persis}}$  &  -0.550 (0.05) & 0.315 (0.05) \\
  $\beta_{\text{dist.neg}}$ & 7.746 (0.57) & 0.285 (0.02)\\
 $\beta_{\text{dist.log }}$  & 0.475 (0.07) & -0.138 (0.04)\\
  $\beta_{\text{water}}$  & -0.132 (0.30)& -1.212 (0.37) \\
 $\beta_{\text{dec}}$  & 0.057 (0.10)& -0.504 (0.10) \\
  $\beta_{\text{meadow}}$ & 0.412 (0.09)& 1.670 (0.09) \\
 $\beta_{\text{road}}$ & -0.247 (1.49)& 1.533 (0.39) \\
   \hline
\end{tabular}
\label{t:resulSSF}
\end{table}

%
%
As was the case in Section~\ref{s:equivalence}, the first state ($k=1$) corresponds to an encamped state while the second state (k=2) corresponds to an exploratory state, i.e., a traveling mode with a moderate significant directional persistence ($\hat{\beta}_{\text{cos.persis}}^{(2)}=0.3147, \ s.e.=0.05 $) and a larger average speed $(\hat{\mathbf{\beta}}_{\text{dist.log }}^{(2)}+1)/ \hat{\mathbf{\beta}}_{\text{dist.neg}}^{(2)}\approx 3.025 $ km per hour.
In the encamped regime, the animal is almost stationary, moving by about 0.1904 km per hour.  The directional persistence parameter $\hat{\beta}_{\text{cos.persis}}^{(1)}$ is strongly negative and significant, which means that the bison tends to move back and forth. In this general model the states are also related to the type of habitat.
In the encamped state ($k=1$) the bison prefers meadows, whereas in the exploratory state ($k=2$), it selectively travels in meadows or roads while avoiding water and deciduous stands.

\begin{figure}[h!]
\centering
\includegraphics[scale=0.6]{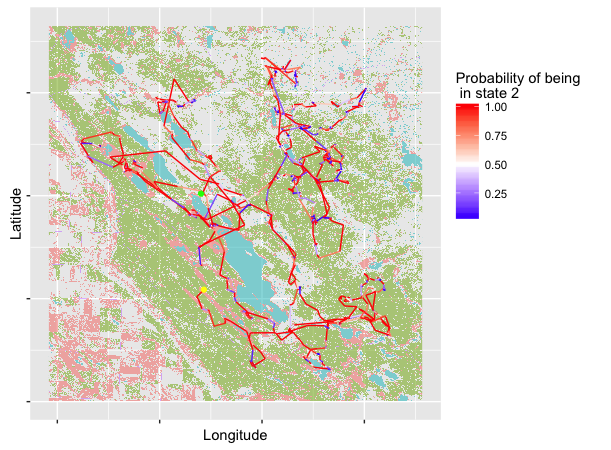}
\caption{Estimated smooth probability for the trajectory of the bison in the landscape presented in Figure~\ref{f:TrajecObsBison} }
\label{f:TrajecObsBisonEstim}
\end{figure}

Table~\ref{t:resulSSF} presents the parameter estimates of the Markov chain model that governs the transitions between states. The stationary distribution of this fitted Markov chain gives a probability of being in state 1 of $\hat q_1/( \hat q_2+\hat q_1)= 0.6053$, suggesting that  the bison was in the travelling regime about  1860 out of $T=3073$ steps.
The model can be used to ``predict'' the state of the bison at time step $t$ using the smooth probabilities $\mathbb{P}(S_{tk}=1 | \mathcal{F}_T; \hat{\mathbb{\theta}}_{\text{MLE}}) $, $t=1,\ldots,T$, $k=1,2$ calculated in the filtering-smoothing part of the E-step of the EM algorithm. These predictions are depicted in Figure~\ref{f:TrajecObsBisonEstim} with a color gradient from red (state 2, $ \mathbb{P}(S_{t1}=1 | \mathcal{F}_T; \hat{\mathbb{\theta}}_{\text{MLE}})=1$) to blue (state 1, $ \mathbb{P}(S_{t2}=1 | \mathcal{F}_T; \hat{\mathbb{\theta}}_{\text{MLE}})=1$).


\section{Conclusion}

This paper proposes a new multi-state version of the SSF model to describe the movement of an animal.
It improves on classical multi-state BCRW modeling by letting two important aspects of animal movement evolve according to multiple behaviors:
a global movement strategy and a local discrete habitat selection; the multi-state BCRW only considered the former.
As such, the proposal generalizes the single-state models proposed by \cite{Dancose2011} or \cite{Avgar2016}.
By using recent techniques for the implementation of the EM algorithm in complex settings, we provide new statistical tools to fit multi-state SSF and identify the hidden behaviors of the animals.

We have proven that the multi-state BCRW of \cite{Nicosia16} can be fitted using the proposed multi-state SSF. This
allows to include explanatory variables that are more general than angles and distances, like for instance the type of land cover, in the analysis. 
Thus gives the method more power to identify the hidden states.
When applied to the analysis of bison movement, the model successfully identified 1) foraging areas and 2) preferred trajectories when the bison moved between foraging areas. First, the strong selection for meadows in encamped mode is consistent with bison spending more time where forage is most abundant. Indeed, bison consume grasses and sedges (plants) that are at least three times more abundant in meadows than forest stands \citep{FortinME2007}. Second, the association between the exploratory mode and habitat features provides valuable information on landscape connectivity. Landscape connectivity involves structural and functional components; structural connectivity depends the physical arrangement of habitat patches, such as their Euclidean distance (\cite{Tischendorf2000}, \cite{Kindlmann2008}), whereas functional connectivity accounts for the movements within the patch network (\cite{Dancose2011},  \cite{Courbin2014}). The exploratory mode model reveals that landscape functional connectivity for bison largely depends on their selective use of roads and meadows for travel, as well as their avoidance of water and deciduous forests relative to mixed and conifer forests. Our study thus demonstrates that multi-state SSF can provide a mechanistic understanding of animal distribution dynamics in heterogeneous landscapes.


In our application, a land animal is followed using the GPS collar technology, which is relatively accurate when compared to other satellite telemetry, such as Argos archival data loggers
that are used to track animal movements in environments like marine systems (\cite{PATTERSON2008}, \cite{Albertsen2015}). Measurement error is therefore not an issue in the present
example, but adding measurement error in the model could be an interesting future development of our method.
There are other possibilities of extension of the methods presented here.
Because animals tend to exhibit heterogeneity in their movement behavior, it would be interesting to carry out the combined analysis of the movement of many individuals using a model featuring random effects.
Defining a multi-state model based on a more complex hidden process could also be potentially interesting, for instance
when trying to model the behavior of an animal over a long period of time (e.g., more than one ``biological season'', see \cite{Basille2012}) where the time homogeneity assumption
becomes questionable.

\section*{Acknowledgements}

The authors are grateful to Marie-Caroline Prima for her help with the bison data and for insightful discussions.  Financial support was provided by a grant from the Fonds de recherche du Qu\'ebec - Nature et technologie to Louis-Paul Rivest, Thierry Duchesne and Daniel Fortin, and a scholarship from the Institut des sciences math\'ematiques awarded to Aur\'elien Nicosia. Field work was supported by the Natural Sciences and Engineering Research Council of Canada (NSERC) and Parks Canada.

\section*{References}

\bibliography{mybib2}
\addtocontents{toc}{\setcounter{tocdepth}{-1}}

\pagebreak
\newpage
\appendix

\section{Equivalence of methods when applied to bison trajectory}
\label{a:A}
\begin{figure}[h!]
 \centering
 \includegraphics[scale=0.8]{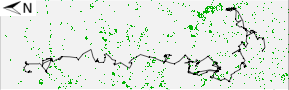}
  \caption{Trajectory of the bison in the Prince Albert National park. Green locations are meadows. Distances in meters.}
  \label{meddow}
 \end{figure}

We now apply the model to data on the movement of bison in Prince Albert National Park, Saskatchewan, Canada. We focus on a single animal wearing a collar recording its location every hour from  January 2012 to February 2012, yielding $T = 992$ consecutive observed locations.
Along with the directional persistence, $\phi_{0,t-1}$, the distance, $d_{0,t}$ and the log-distance, $\ln d_{0,t}$, other covariates based on the type of landscape can be considered in the analysis,
but among the latter only the direction to the closest meadow is kept, as suggested by \cite{Dancose2011}. Figure~\ref{meddow} displays the trajectory of the bison with respect to the meadows available in the park.
%

We first fit the multi-state general random walk model with observed direction depending on the vector (3.1), as
\begin{equation} \tag{A.1}
\mathbf{V}_t^{(k)}=\kappa_0^{(k)} \left(\begin{array}{c}
\cos(\phi_{0,t-1})\\
\sin(\phi_{0,t-1})
\end{array}  \right)+ \kappa_1^{(k)} \left(\begin{array}{c}
\cos(\theta_{t})\\
\sin(\theta_{t})
\end{array} \ \right) , t=1,\ldots,992,
\end{equation}
where the angle $\theta_{t}$ denotes the direction from the current location $P_t$ of the bison at time step $t$ to the closest meadow.
We then fit the multi-state SSF model with $J=500$ control distances sampled uniformly over $[0,M]$ for each observed location, where $M=1$km.
 This value of $M$ is the maximum of the two quantiles of order $0.9999$ of the two distance distributions obtained with the BCRW.
 Finally we fit the SSF model with $J=500$ control locations again, but we sample the control distances using an exponential distribution with rate 1.
 The SSF equivalent to the BCRW given by (3.3) has covariate vector
$$
\mathbf{x}_{ti}=\left(\log(h_{ti}),-h_{ti},\cos(\phi_{ti}-\phi_{t-1,0}),\cos(\phi_{ti}-\theta_{t})\right)^\top.
$$
The results are summarized in Table~\ref{t:resultCar} and they confirm the theoretical results of Section 3.1 that all three models are equivalent for $J$ large enough.

 \begin{table}[h!]
 \caption{Estimators and standard errors of the parameters obtained when fitting the BCRW, the SSF with uniform sampling of the control distance and $J=500$ and
 the SSF with parametric unit exponential sampling of the control distances and $J=500$ to the bison data. (The distances were divided by 100 to make the
 estimates of the parameters of the distance distributions easier to report.) }
 \centering
 {\renewcommand{\arraystretch}{1.5} 
{\setlength{\tabcolsep}{0.4cm} 
\begin{tabular}{|c c c|| c c | c c  |}
\hline
\multicolumn{3}{|c||}{Multi-states BCRW} & \multicolumn{4}{|c|}{Multi-states SSF} \\
& \multicolumn{2}{c||}{}& \multicolumn{2}{c}{uniform}& \multicolumn{2}{c|}{ Corrected parametric} \\

Parameters& $\hat{ \theta}$ & Sd$(\hat{\theta})$& $\hat{ \theta}$ & Sd$(\hat{\theta})$& $\hat{ \theta}$ & Sd$(\hat{\theta})$  \\
\hline
\hline
$q_1$ &  0.6941 & 0.0364&0.7023& 0.0364 &0.6912 &0.0368  \\
$q_2$ &  0.1765 &0.0207 & 0.1781 & 0.0201 &0.1744 &0.0203\\
 &&&&&& \\
$\kappa_0^{(1)}=  \beta_{3}^{(1)} $ & 0.2605 & 0.0878& 0.2409& 0.0876& 0.2542& 0.0879\\

$\kappa_1^{(1)}=\beta_{4}^{(1)}$ & 0.1559 & 0.0865&  0.1588& 0.0868& 0.1533&0.0870 \\

$\lambda_1^{(1)}-1=\beta_{1}^{(1)}$ & 0.0684 &   0.1161&0.0253& 0.1201&0.0843 &0.1216\\

$1/\lambda_2^{(1)}=\beta_{2}^{(1)}$ &0.5880 & 0.0564 &0.5654& 0.0512 &0.5847 &0.0588 \\

 &&&&&& \\

$\kappa_0^{(2)}=\beta_{3}^{(2)}$ & -0.4535& 0.0670& -0.4822& 0.0665&-0.4494 &0.0663 \\

$\kappa_1^{(2)}=\beta_{4}^{(2)}$ & 0.1036 &  0.0640&0.0869 & 0.0639& 0.1025&0.0637\\

$\lambda_1^{(2)}-1=\beta_{1}^{(2)}$ &  0.3747 & 0.0813&0.3684 & 0.0807 &0.3684 &0.0815  \\

$1/\lambda_2^{(1)}=\beta_{2}^{(2)}$ & 6.90359 &  0.6531&    6.9241&0.6499 & 6.8118&0.6294 \\

\hline
\end{tabular} }}
\label{t:resultCar}
 \end{table}

The analysis on a real bison trajectory shows that the proposed multi-state SSF is equivalent to the general multi-state random walk model. Moreover one can find the classical encamped and explanatory states exhibited by the bison (see \cite{Langrock2012} and \cite{Nicosia16} for more details.)

\section{Proofs of the equivalence between the proposed multi-state SSF model and the multi-state BCRW model}
\label{a:B}

This appendix gives the proofs of the equivalence between the proposed multi-state SSF model and the multi-state BCRW model (section 3).

\subsection{Equivalence with uniform sampling of the control distances} \label{a:prop1}

For each state $k=1,\ldots,K$, let $M_{\epsilon^{(k)}}$ denote the positive constant defined such that:
$$
\int_{0}^{M_{\epsilon^{(k)}}}  g_k(x; \eta^{(k)}) dx \geq 1- \epsilon^{(k)},
$$
for $\epsilon^{(k)}$ small enough, i.e., $M_{\epsilon^{(k)}}$ is the quantile of order $\epsilon^{(k)}$ of the distribution of the distance in the state $k$. Let also $M \geq \max_{k=1\ldots,K} M_{\epsilon^{(k)}}$ defined for $ \max_{k=1\ldots,K} \epsilon^{(k)}$ small enough.

Using the specification of the multi-state SSF model, we have that $p_t^{(k)}$, the conditional likelihood at time $t$ given that the animal is in state $k$, is proportional to
\begin{align*}
p_t^{(k)} &= \frac{\exp \left\lbrace\mathbf{x}_{0t}^\top \beta^{(k)}+\ln b(h_{0t})\right\rbrace}{\sum_{j = 0}^J \exp \left(\mathbf{x}_{jt}^\top \beta^{(k)}+\ln b(h_{0t}) \right)} \\
& \propto  \frac{J b(h_{0t})\exp \left(\mathbf{x}_{0t}^T {\beta}^{(k)} \right)}{ b(h_{0t})\exp\left(\mathbf{x}_{0t}^T {\beta}^{(k)} \right)+ \sum_{j = 1}^J b(h_{jt})\exp\left(\mathbf{x}_{jt}^T {\beta}^{(k)} \right)} \\
&= \frac{b(h_{0t})\exp \left(\mathbf{x}_{0t}^T {\beta}^{(k)} \right)}{\frac{b(h_{0t})\exp \left(\mathbf{x}_{0t}^T {\beta}^{(k)} \right)}{J}+ \frac{1}{J}\sum_{j = 1}^J b(h_{jt}) \exp\left(\mathbf{x}_{jt}^T {\beta}^{(k)} \right)}.
\end{align*}
When $J$ is large, the term in the denominator involving $\mathbf{x}_{0t}^T$ is negligible, and by the law of large numbers as $J$ goes to infinity $\frac{1}{J} \sum_{j = 1}^J  b(h_{jt}) \exp\left(\mathbf{x}_{jt}^T {\beta}^{(k)} \right)$ converges to
$$
\mathbb{E}_{\Theta} \left[ e^{ \kappa_0^{(k)} \cos(\Theta-\phi_{0,t-1}) +\sum_{i=1}^p \kappa_i^{(k)} \cos(\Theta-\theta_{it})} \right] \times \mathbb{E}_{D} \left[ b(D)e^{\eta^{(k)} T(D)}  \right],
$$
where $\Theta \sim \mathcal{U}(\mathcal{S}^1)$ and $D\sim \mathcal{U}([0,M])$ because the control locations are sampled uniformly over the disc of radius $M$. We can evaluate these expectations:
\begin{align} 
\mathbb{E}_{\Theta} \left[ e^{ \kappa_0^{(k)} \cos(\Theta-\phi_{0,t-1}) +\sum_{i=1}^p \kappa_i^{(k)} \cos(\Theta-\theta_{it})} \right] &=\int_{\mathbb{R}} e^{ \kappa_0^{(k)} \cos(\theta-\phi_{0,t-1}) +\sum_{i=1}^p \kappa_i^{(k)} \cos(\theta-\theta_{it})} d F_{\Theta}(\theta) \nonumber \\
&=\frac{1}{2 \pi} \int_{-\pi}^{\pi} e^{ \kappa_0^{(k)} \cos(\theta-\phi_{0,t-1}) +\sum_{i=1}^p \kappa_i^{(k)} \cos(\theta-\theta_{it})} d \theta \nonumber \\
&= I_0(\ell_t^{(k)}) \tag{B.1}
\end{align}
and
\begin{align} 
\mathbb{E}_{D} \left[ b(D)e^{\eta^{(k)} T(D)}  \right]  &= \int_{\mathbb{R}} b(t)e^{\eta^{(k)} T(t)} dF_D(t)  \nonumber\\
&=\frac{1}{M}\int_0^M b(t)e^{\eta^{(k)} T(t)} dt  \nonumber\\
&\propto \exp A(\eta^{(k)}) \tag{B.2} ,
\end{align}
since
\begin{align*} 
\int_0^{M} b(t) \exp \lbrace \eta^{(k)} T(t) \rbrace d t &=\int_0^{M_{\epsilon^{(k)}}} b(t) \exp \lbrace \eta^{(k)} T(t) \rbrace d t +  \int_{M_{\epsilon^{(k)}}}^M b(t) \exp \lbrace \eta^{(k)} T(t) \rbrace d t  \\
& \underset{ \epsilon^{(k)} \rightarrow 0}{\approx}   \exp A(\eta^{(k)}) + 0.
\end{align*}

Finally when $J$ becomes large and combining (A.1) and (B.2) in:
\begin{align}
p_t^{(k)} &\propto \frac{e^{\mathbf{x}_{t0}^T {\beta}^{(k)}}}{ \mathbb{E}_{\Theta} \left[ e^{ \kappa_0^{(k)} \cos(\Theta-\phi_{0,t-1}) +\sum_{i=1}^p \kappa_i^{(k)} \cos(\Theta-\theta_{it})} \right] \times \mathbb{E}_{D} \left[ b(D)e^{\eta^{(k)} T(D)}  \right]} \nonumber \\
&= \frac{b(d_{0t}) e^{\eta^{(k)} T(d_{0t})}}{ \exp A(\eta^{(k)})} \times \frac{e^{ \kappa_0^{(k)} \cos(\phi_{0t}-\phi_{0,t-1}) +\sum_{i=1}^p \kappa_i^{(k)} \cos(\phi_{0t}-\theta_{it})}}{ I_0(\ell_t^{(k)})} \nonumber \\
& \propto  g_k(d_{0t};\eta^{(k)}) \times f_k(\phi_{0t}|\mathcal{F}_{t-1}^o)  \tag{B.3},
\end{align}
where $f_k$ is the density of the von Mises distribution. This completes the proof as (B.3) is the conditional likelihood at time step $t$ and state $k$ of the general hidden random walk model of \cite{Nicosia16}.

\subsection{Equivalence with parametric sampling of the control distances}

The proof is quite similar to the proof when the control distances are uniformly sampled. Due to the law of large numbers the conditional probability in the step selection function model is proportional to
$$
p_t^{(k)} \propto  \frac{e^{\eta^{(k)} T(d_t)}}{ \int_{\mathbb{R}} e^{\eta^{(k)} T(t)} dF_D(t) } \times f_k(\phi_{0t}|\mathcal{F}_{t-1}^o),
$$
where $f_k$ is the density of the von Mises distribution. The integral at the denominator is taken over the distribution of $D$, i.e. from (3.2) with parameters $\tilde{\eta}$ and can be evaluated as follows:
\begin{align*}
 \int_{\mathbb{R}} e^{\eta^{(k)} T(t)} dF_D(t)&= \int_0^{\infty} b(t) e^{\eta^{(k)} T(t)}  e^{\tilde{\eta}^{(k)} T(t)-A(\tilde{\eta}^{(k)} )}dt \\
&= e^{-A(\tilde{\eta}^{(k)} )} \int_0^{\infty} b(t) e^{(\eta^{(k)}+\tilde{\eta}^{(k)} ) T(t)} dt \\
&= e^{-A(\tilde{\eta}^{(k)} )} e^{A(\eta^{(k)}+\tilde{\eta}^{(k)})}.
\end{align*}
Thus the conditional likelihood of the discrete choice of the animal at time step $t$ and state $k$ is
$$
p_t^{(k)} \propto  \frac{b(d_{0t}) e^{(\eta^{(k)}+\tilde{\eta}^{(k)}) T(d_{0t})}}{ e^{A(\eta^{(k)}+\tilde{\eta}^{(k)}) }} \times f_k(\phi_{0t}|\mathcal{F}_{t-1}^o),
$$
which ends the proof.

\section{Numerical implementation of maximum likelihood Estimation}
\label{a:C}

The EM algorithm is generally used for the maximization of likelihood functions when data are missing or unobserved.
The EM algorithm only requires evaluation of the complete data log-likelihood function, which in our case is easily derived from (2.3):
\begin{eqnarray*}
\log \text{L}_{\text{complete}}({\theta};\mathcal{F}_T^c) &=&  \sum_{t=1}^T \sum_{h=1}^K \sum_{k=1}^K S_{h,t-1}S_{k,t} \log \pi_{hk} \\
&+& \sum_{t=1}^T \sum_{k=1}^K S_{kt} \log p_t^{(k)}  .\\
\end{eqnarray*}
The EM algorithm consists of iterating an expectation (E) and a maximization (M) step. Let us denote by
$\hat{{\theta}}_s$ the value of the estimate of ${\theta}$ after the $s$-th iteration of the algorithm. Then the $(s+1)$-th iteration of the algorithm starts with one application of the E-step, which evaluates the expectation of $\log \text{L}_{complete}$ with respect to the conditional distribution of the missing values given the observed data, as follows:
\begin{eqnarray*}
Q({\theta}|\hat{{\theta}}_s )&=& \mathbb{E}_{S_{0:T}}\left[\log \text{L}_{\text{complete}}({\theta};\mathcal{F}_T^c)|\mathcal{F}_{T}^o,\hat{{\theta}}_s \right] \nonumber \\
&=& \sum_{t=1}^T \sum_{h=1}^K \sum_{k=1}^K \mathbb{E} (S_{h,t-1}S_{k,t}|\mathcal{F}_{T}^o,\hat{{\theta}}_s) \log \pi_{hk}\\
&+& \sum_{t=1}^T \sum_{k=1}^K \mathbb{E}( S_{kt}|\mathcal{F}_{T}^o,\hat{{\theta}}_s) \log p_t^{(k)} .  \\
\end{eqnarray*}
Then the value of $\hat{{\theta}}_{s+1}$ is calculated in the M-step as the value of ${\theta}$ that maximizes
 $Q({\theta}|\hat{{\theta}}_s )$.

\subsection*{• E step}
The function $Q(.|\hat{{\theta}}_s)$ involves two conditional expectations, $\mathbb{E}( S_{kt}|\mathcal{F}_{T}^o,\hat{{\theta}}_s) $ and $\mathbb{E} (S_{h,t-1}S_{k,t}|\mathcal{F}_{T}^o,\hat{{\theta}}_s)  $. These can be efficiently computed by a forward-backward (filtering-smoothing) algorithm for Markov chains, see Appendix D. The filtering-smoothing algorithm starts from the initial time $t=0$ and computes the ``filtering" probabilities $\mathbb{P}(S_t|\mathcal{F}_t^o)$ by using predictive probabilities $\mathbb{P}(S_t|\mathcal{F}_{t-1}^o)$ (going forward in time). The last filtering probability $\mathbb{P}(S_T|\mathcal{F}_T^o)$ is then used to compute the ``smoothing" probabilities $\mathbb{P}(S_t|\mathcal{F}_T^o)$ using Bayes theorem (going backward in time).
Details of this implementation of the E-step in a context of random walk model is given in the appendix of  \cite{Nicosia16}.

\subsection*{• M step}
 For the M-step, we see that $Q({\theta}|\hat{{\theta}}_s ) $ is a sum of three functions that depend on different sets of parameters and can thus be maximized separately:

\begin{itemize}

\item When the latent states follow a Markov process, there is a closed form expression for the maximizer of the hidden process part,
$$
\hat{\pi}_{hk}^{(s+1)}=\frac{\sum_{t=1}^T  \mathbb{E} (S_{h,t-1}S_{k,t}|\mathcal{F}_{T}^o,\hat{{\theta}}_s)}{\sum_{t=1}^T \mathbb{E} (S_{h,t-1}|\mathcal{F}_{T}^o,\hat{{\theta}}_s)}, h,k=1,\ldots,K.
$$

\item Since $p_t^{(k)}$ has a conditional logistic regression form, the log-likelihood for the observed choice can be maximized with respect to ${\beta}^{(k)}$ using a weighted maximum likelihood procedure (e.g. the function \texttt{coxph} with \texttt{weigths} in the \textsf{survival} \texttt{R} package, see \cite{survival-package} and \cite{survival-book}).

\end{itemize}
%
%
\subsection*{Sampling Distributions}

Quantities that are usually required for inference such as the value of the maximized log-likelihood for the observed data or an estimation of the variance matrix of $\hat{\mathbf{\theta}}$ are not
directly computed when using the EM-algorithm.
The filtering-smoothing algorithm is used to evaluate the observed data likelihood  (2.3). Moreover at each time $t$, one can evaluate the probability that the animal is in state $k$ using the value of $\mathbb{E}(S_{kt}|\mathcal{F}_T)$ in the ``smoothing'' part of the Filtering-Smoothing algorithm.
 Because we are able to compute $\log \text{L}(\hat{\mathbf{\theta}}_{\text{MLE}})$, we can numerically approximate the negative of its hessian matrix,
whose inverse, denoted ${v}$, is the usual estimate of the variance matrix of the maximum likelihood estimators. A numerical approximation of the Hessian matrix is available under most software implementations
of the Broyden--Fletcher--Goldfarb--Shanno (BFGS) algorithm  \citep{avriel2003nonlinear};  in the data analysis section
 we use the one provided in the R function \texttt{optim}.

%

\section{Filtering-Smoothing Algorithm}
\label{a:D}

In the E-step of the ($s$+1)-th iteration of the EM algorithm we have to compute two posterior expectations involving the hidden $S_{kt}$, $k=1,\ldots,K$, $t=0,\ldots,T$, conditionally on the observed data  $\mathcal{F}_{T}^o$:
\begin{align*}\tag{D.1}
\label{E1}
\mathbb{E}( S_{kt}|\mathcal{F}_{T}^o,\hat{{\theta}}_s)&=\mathbb{P}( S_{kt}=1|\mathcal{F}_{T}^o,\hat{{\theta}}_s)
\end{align*}
\begin{equation} \tag{D.2}
\mathbb{E} (S_{h,t-1} S_{k,t}|\mathcal{F}_{T}^o,\hat{{\theta}}_s) =\mathbb{P}(S_{t-1,h}=1|S_{tk}=1,\mathcal{F}_{T}^o,\hat{{\theta}}_s)\mathbb{P}(S_{tk}=1 |\mathcal{F}_{T}^o,\hat{{\theta}}_s), \label{E3}
\end{equation}
where $\hat{{\theta}}_s$ is the maximized vector of parameters after the $s$-th step of the EM algorithm. The first probability on the RHS of (\ref{E3}) can be computed with Bayes' theorem because, as we can see from Figure 1.
, $S_{t-1}$ is independent of the observed data from time $t$ to $T$ (i.e. $\lbrace \mathcal{F}_{t+s}^o \rbrace_{s \geq 0} \setminus \mathcal{F}_{t-1}^o$)  given $S_t$ and $\mathcal{F}_{t-1}^o $ :
\begin{eqnarray*}
\mathbb{P}(S_{t-1,h}=1|S_{tk}=1,\mathcal{F}_{T}^o,\hat{{\theta}}_s)=\frac{\hat{\pi}_{hk}^{(s)} \mathbb{P}(S_{t-1,h}=1 |\mathcal{F}_{t-1}^o,\hat{{\theta}}_s)}{\sum_{j=1}^K \hat{\pi}_{jk}^{(s)} \mathbb{P}(S_{t-1,j}=1 |\mathcal{F}_{t-1}^o,\hat{{\theta}}_s) } ,\ k=1,\ldots,K,\ t=0,\ldots,T.
\end{eqnarray*}
Finally, to compute the remaining conditional probabilities in the posterior expectations \eqref{E1} and \eqref{E3}, we adapt the filtering-smoothing algorithm of \cite{Sylvia13}.
\\
\rule{\linewidth}{.5pt}
\\
\textsc{Filtering-smoothing algorithm to implement the E-step \\ of the  $(s+1)$-th iteration of the EM algorithm.}
\\
\rule{\linewidth}{.5pt}
\\
\begin{itemize}
\item[\textbf{Filter}] Compute $\mathbb{P}(S_{tl}=1|\mathcal{F}_{t}^o,\hat{{\theta}}_s)$, for every $l=1,\dots,K$ :
$$
\mathbb{P}(S_{tl}=1|\mathcal{F}_{t}^o,\hat{{\theta}}_s)=\frac{  p_t^{(l)} \mathbb{P}(S_{tl}=1|\mathcal{F}_{t-1}^o,\hat{{\theta}}_s)}{\sum_{k=1}^K p_t^{(k)} \mathbb{P}(S_{tk}=1|\mathcal{F}_{t-1}^o,\hat{{\theta}}_s)}
$$
where $\mathbb{P}(S_{1l}=1|\mathcal{F}_{0}^o,\hat{{\theta}}_s)=\sum_{k=1}^K \hat{\pi}_{kl}^{(s)}({\pi}_0)_k$ \\ and
\begin{equation*}
\label{a:Predictive}
\mathbb{P}(S_{tl}=1|\mathcal{F}_{t-1}^o,\hat{{\theta}}_s)=\sum_{k=1}^K \hat{\pi}_{kl}^{(s)}\mathbb{P}(S_{t-1,k}=1|\mathcal{F}_{t-1}^o,\hat{{\theta}}_s)
\end{equation*}
for $t=2,\dots,T.$
\item[\textbf{Smooth}] Compute $\mathbb{P}(S_{tl}=1|\mathcal{F}_{T}^o,\hat{{\theta}}_s)$, for every $l=1,\dots,K$:
\begin{itemize}
\item[\textsf{S-step 1}] For $t=T$, set $\mathbb{P}(S_{Tl}=1|\mathcal{F}_{T}^o,\hat{{\theta}}_s)$, the conditional probability computed
at the last filtering step.

\item[\textsf{S-step 2}] Recursion: For $t=T-1,\ldots,0$, compute:
\begin{equation*}
\label{a:smooth}
\mathbb{P}(S_{tl}=1|\mathcal{F}_{T}^o,\hat{{\theta}}_s)=\sum_{k=1}^K \frac{\hat{\pi}_{lk}^{(s)}\mathbb{P}(S_{tl}=1|\mathcal{F}_{t}^o,\hat{{\theta}}_s) \mathbb{P}(S_{t+1,k}=1|\mathcal{F}_{T}^o,\hat{{\theta}}_s)}{\sum_{j=1}^K \hat{\pi}_{jk}^{(s)} \mathbb{P}(S_{tj}=1|\mathcal{F}_{t}^o,\hat{{\theta}}_s)}.
\end{equation*}

\end{itemize}

\end{itemize}

\rule{\linewidth}{.5pt}

\end{document}